\documentclass{ws-procs975x65}

\newcommand{\be}{\begin{equation}}
\newcommand{\ee}{\end{equation}}
\newcommand{\bea}{\begin{eqnarray}}
\newcommand{\eea}{\end{eqnarray}}

\begin{document}

\title{Adiabatic regularization for spin $1/2$ fields and the renormalized stress-energy tensor}
\author{Adrian del Rio$^*$ and Jose Navarro-Salas$^{\dagger}$}

\address{Department of Theoretical  Physics, IFIC. University of Valencia --- CSIC.\\
Valencia, 46020, Spain\\
$^*$ adrian.rio@uv.es, $^{\dagger}$ jnavarro@ific.uv.es}

\begin{abstract}
In this article we briefly discuss the adiabatic renormalization program for spin 1/2 fields in expanding universes. We introduce the method and provide explicit expressions for the renormalized vacuum expectation value of the stress-energy tensor. Then, we discuss its application to some cosmological scenario of physical interest. We end up sketching out the proof that adiabatic  and DeWitt-Schwinger point-splitting schemes provide the same renormalized expectation values of the stress-energy tensor for Dirac fields. 
\end{abstract}

\bodymatter


\section{Introduction and Motivation}

One of the most important consequences of combining quantum theory with general relativity is the phenomenon of gravitational particle creation, as first discovered by Parker [\refcite{parker}] (see also the reviews [\refcite{parker-toms}, \refcite{birrell-davies}]). The generation and amplification of quantum field fluctuations is inevitable during the expansion of the universe, and hence the creation of quanta. 
In order to create a significant amount of quantum fluctuations, we need rapid expansions, like those expected to happen in the very early universe [\refcite{inflation}]. This is essentially the mechanism driving the generation of primordial inhomogeneities observed in the large-scale structure of the universe and in the temperature distribution of the cosmic microwave background [\refcite{inflation2}]. 

In the quantum theory, the gravitationally produced perturbations contribute to the energy density and pressure of the field with new ultraviolet (UV) divergences, not present in the quantization of free fields in Minkowski space-time. As a consequence, one needs to use a self-consistent regularization and renormalization scheme in curved space-time to subtract these divergences properly. One of the most useful and powerful schemes in cosmological scenarios is the adiabatic regularization method [\refcite{parker-fulling}]. In the case of scalar fields, this method is based on a WKB-type expansion of the field modes, which allows  to identify the divergent terms of the tensor unequivocally and to subtract them from the bare expressions. 

One of the main issues with the renormalization program in curved space-time is that these methods have been mainly developed for free scalar bosons, and less work has been done for other fields. In particular, an adiabatic regularization method for spin-$1/2$ fields in an expanding universe was missing until very recently [\refcite{landete}, \refcite{papers}]  (see also [\refcite{ghosh}]). In this paper, we sum up the main results of the adiabatic regularization method, and provide general and explicit expressions for the renormalized stress-energy tensor of a spin-$1/2$ field in a Friedmann-Lemaitre-Robertson-Walker (FLRW) universe. This result is written in terms of UV-convergent momentum integrals involving the field modes. This is a necessary and unavoidable step to prepare the method to be used for numerical computations in  cosmology.

\section{Quantized Spin $1/2$ Fields and the Adiabatic Expansion}

 The equation of motion for a spin-1/2 field $\psi(x)$ of mass $m$ in curved spacetime is given by the Dirac equation:
$ (i{\gamma}^{a}e^{\mu}_{a}\nabla_{\mu}-m)\psi(x)=0 \ , \label{1}$
where ${\gamma}^{a}$ are the Dirac matrices satisfying the Clifford algebra $\{{\gamma}^{a},{\gamma}^{b}\}=2\eta^{ab}$;  $\nabla_{\mu}$ is the covariant derivative operator; and $e^{\mu}_{a}$ is the vierbein basis, a metric preserving isomorphism between Minkowski space (denoted with indices $a$) and the tangent space at each spacetime point (denoted with space-time indices $\mu$), $g_{\mu\nu}=e_{\mu}^a e_{\nu}^b\eta_{ab}$.

In a spatially flat FLRW universe,  $ds^2=dt^2-a^2(t)d\vec{x}^2$, one can take advantage of the spatial symmetries (homogeneity and isotropy) in order to simplify the equations of motion, by expanding the Dirac field in a complete set of Fourier modes,
\bea
\psi(x)=\int d^3\vec{k} \sum_{\lambda}\left[B_{\vec{k} \lambda}u_{\vec{k} \lambda}(x)+D_{\vec{k} \lambda}^{\dagger}v_{\vec{k} \lambda}(x) \right] \ .  \label{4c}
\eea 
Symmetries of  space-time allow to discompose the spatial dependence of the modes (whose solution of the Dirac equation is simply a plane wave), from the temporal one.
By extending the quantization procedure in Minkowski space  one can construct, for a given mode $\vec k$,  two independent spinor solutions  as 
 \bea
u_{\vec{k}\lambda}(x)=u_{\vec{k}\lambda}(t)e^{i \vec{k} \cdot \vec{x}}=\frac{e^{i \vec{k} \cdot \vec{x}}}{\sqrt{(2\pi)^3 a^3}}
\begin{pmatrix}
h_k^I(t) \xi_{\lambda}(\vec{k})  \\
h_k^{II}(t) \frac{\vec{\sigma}\cdot \vec{k}}{k} \xi_{\lambda}(\vec{k})   
\end{pmatrix},  \label{3}
\eea 
 where   $\xi_{\lambda}$ is a constant and normalized two-component spinor $\xi_{\lambda}^{\dagger} \xi_{\lambda '}=\delta_{\lambda'\lambda}$.  It is convenient to use helicity eigenstates  $\xi_{\lambda}(\vec k)$, which follow the property  $\frac{\vec{\sigma}\vec{k}}{2k}\xi_{\lambda}(\vec{k})=(\lambda/2) \xi_{\lambda}(\vec{k})$, where $\lambda/2= {\pm}1/2$ represent the eigenvalues for the helicity.

From this space of solutions one constructs a Hilbert space by defining the Dirac inner product:
 $
(\psi_1,\psi_2)=\int d^3x a^3 \psi_1^{\dagger} \psi_2 \ ,    \label{3e}
$
 and requiring modes to be normalized according to $(u_{\vec{k}\lambda},u_{\vec{k}\,'\lambda'})=(v_{\vec{k}\lambda},v_{\vec{k}\,'\lambda'})=\delta_{\lambda\lambda'}\delta^{(3)}(\vec{k}-\vec{k}\,')$ and $(u_{\vec{k}\lambda},v_{\vec{k}\,'\lambda'})=0$, which ensures the standard anticommutation relations for annihilation/creation operators $B_{\vec{k} \lambda}$ and $ D_{\vec{k} \lambda}$.

Following this decomposition (\ref{3}),  only  the temporal dependence of the modes remains to be solved. The general solution to the differential Dirac equation admits free parameters  reflecting the so-called ambiguity of the vacuum state in the quantum theory, which has to be fixed by additional physical considerations. In order to deal with UV divergences, though, only the (geometric) asymptotic structure in the high frequency $k$  (flatness limit) is needed, which is shared by all solutions of the differential equation of motion. For expanding universes, this asymptotic behaviour is translated to the time evolution and is called adiabatic. 
The adiabatic regularization method for spin-1/2 fields, introduced in [\refcite{landete}], is based on the following ansatz for the field modes
\be h^{I}_{{k}}(t) \sim \sqrt{\frac{\omega + m}{2\omega}} e^{-i \int^{t'} \Omega (t') dt'} F(t) \,\,\,\,\,\, \ ,\,\,\,\,\,\, 
h^{II}_{{k}}(t) \sim \sqrt{\frac{\omega - m}{2\omega}} e^{-i \int^{t'} \Omega (t') dt'} G(t)\ , \label{fermion-ansatz} \ee
where $\omega \equiv \sqrt{(k/a(t))^2 + m^2}$ is the frequency of the mode and the  time-dependent functions $\Omega (t)$, $F(t)$ and $G(t)$ are expanded in power series ('adiabatically') as
\bea \Omega (t) &=& \omega + \omega^{(1)} + \omega^{(2)} + \omega^{(3)} + \omega^{(4)} + \dots \ ,  \nonumber \\
F (t) &=& 1 + F^{(1)} + F^{(2)} + F^{(3)} + F^{(4)} + \dots \ ,  \nonumber \\
G (t) &=& 1 + G^{(1)} + G^{(2)} + G^{(3)} + G^{(4)} + \dots \ . \label{adifexp}\eea
Here, $\omega^{(n)}$, $F^{(n)}$ and $G^{(n)}$ are functions of adiabatic order $n$, which means that they contain $n$ derivatives of the scale factor. In the expansions above, we impose $F^{(0)} = G^{(0)} \equiv 1$, and $\omega^{(0)} \equiv \omega$, to recover the Minkowskian solutions in the adiabatic limit. 
Using the Dirac equation the system can be solved by iteration, and explicit solutions can be obtained in terms of lower adiabatic orders  [\refcite{papers}].

\section{Renormalization of the Stress-Energy Tensor}

We can employ this adiabatic program to obtain an explicit expression for the renormalized stress-energy tensor for Dirac fields in expanding universes.
The classical stress-energy tensor for a Dirac field in curved space-time is given by
\bea
T_{\mu\nu}=\frac{i}{2}\left[\bar{\psi}\, {\gamma}_{a} e^a_{(\mu} \nabla_{\nu)}\psi-\bar{\psi}\  \overleftarrow{\nabla}_{(\nu}\, e^a_{\mu)}{\gamma}_{a}\psi \right]  \ .  \label{10}
\eea
Due to spatial symmetries we only have to consider two independent components. Namely, the energy density (related with the 00-component) and the pressure (related with the ii-component). Performing a detailed calculation, the vacuum expectation values can be seen to give [\refcite{papers}]
\bea
\left< T_{00}\right>=\frac{1}{2\pi^2 a^3}\int_{0}^{\infty} dk k^2 \rho_k  \ , \hspace{1cm}  \rho_k (t) \equiv 2i \left( h_k^{I} \frac{\partial h_k^{I*}}{\partial t} - h_k^{I*} \frac{\partial h_k^{I}}{\partial t} \right) \ , \label{12}
\eea
for the energy density, and
\bea
\left< T_{ii}\right>=\frac{1}{2\pi^2 a}\int_{0}^{\infty} dk k^2 p_k  \ ,  \hspace{1cm}  p_k(t)\equiv-\frac{2k }{3a}[h_k^{I} h_k^{II*}+h_k^{I*} h_k^{II}]  \ , \label{13c}
\eea 
for the pressure.
As can be seen from the above results, both energy density and pressure can be written in terms of the field modes, so that the adiabatic program can be implemented naturally. The stress-energy tensor contains different ultraviolet divergences, and so, we must expand the integrands adiabatically  $\rho_k = \rho_k^{(0)} + \rho_k^{(1)} + \rho_k^{(2)} + \dots$,  and remove from (\ref{12})-(\ref{13c}) enough terms in order to yield a UV-finite quantity, 
\bea
\langle T_{00} \rangle_{ren} & \equiv &    \langle T_{00} \rangle - \langle T_{00} \rangle_{Ad}=      \frac{1}{2 \pi^2 a^3} \int_0^{\infty} dk k^2 ( \rho_k - \rho_k^{(0)} - \rho_k^{(2)} - \rho_k^{(4)} ) \ ,\\
\left< T_{ii}\right>_{ren}& \equiv &  \left< T_{ii}\right> - \left< T_{ii}\right>_{Ad}= \frac{1}{2\pi^2 a}\int_{0}^{\infty} dk k^2 \left[p_k-p_k^{(0)} -p_k^{(2)}-p_k^{(4)} \right] \ .  \label{17b}
\eea
The counterterms needed to yield a finite quantity of the stress-energy tensor are geometric objects, and thus can be reabsorbed in the coupling constants of the gravitational action. Explicit final results can be looked up in [\refcite{papers}]. We note in particular that the adiabatic program recovers  the standard trace anomaly for a Dirac field in a curved background, $\left< T_{\mu}^{\mu}\right>_{ren}$ ,  and it respects  stress-energy conservation $\nabla^{\mu}\left< T_{\mu\nu}\right>_{ren}=0$ and general covariance, order by order [\refcite{landete}, \refcite{papers}].

As an example to illustrate the power of the adiabatic method, we can analyze the corresponding implications for a radiation dominated universe, where $a(t)\sim t^{1/2}$. This is a particularly nice case, since the Dirac equation for the modes can be solved analitically. The general solution is given by a linear combination of Whittaker $W$ functions [$z=z(t)$, $\kappa=\kappa(k)$, and $N$ is just a normalization factor]
\bea
h_k^{I}= E_k \left ( N \frac{W_{\kappa,\frac{1}{4}}(z)}{\sqrt{a(t)}} \right )    
+ F_k \left ( N \frac{k}{2m a(t)^{3/2} }\left[W_{\kappa,\frac{1}{4}}(z)+\left(\kappa-\frac{3}{4}\right)W_{\kappa-1,\frac{1}{4}}(z) \right]   \right )^* \ , \label{34}
\eea
where the two arbitrary parameters $E_k$ and $F_k$ reflect the ambiguity in the choice of a vacuum state. Normalization implies $|E_k|^2+|F_k|^2=1$. The adiabatic condition  for large momenta  requires  $E_k \sim 1$ and $F_k \sim 0$ as $k \to \infty$. Moreover, a detailed analysis of the asymptotic properties of the Whittaker functions provides the necessary and sufficient condition for the renormalizability of the stress-energy tensor vacuum expectation value  [\refcite{papers}]. It is
$ \label{renor} |E_k|^2 - |F_k|^2 = 1 + O(k^{-5}) \ . $

Unlike Minkowski or de Sitter spaces, the absence of a maximal group of symmetries for the radiation-dominated background  does not fix a natural  preferred vacuum state. However, the early and late-time behaviors ($t << m^{-1}$ and $t>> m^{-1}$, respectively) of the renormalized stress-energy tensor can be obtained generically, and agree with the forms assumed by classical cosmology. As detailed in [\refcite{papers}], we have that, as time evolves and reaches the regime $ t >> m^{-1}$, the renormalized energy density takes the form of cold matter 
\bea
 \left<T_{00} \right>_{ren} \sim \frac{\rho_{0m}}{a^3} \ , \hspace{0.5cm} \rho_{0m} = \frac{m}{ \pi^2}\int_0^{\infty} dk k^2 \left[1-(|E_k|^2-|F_k|^2) \right] \geq 0 \ . 
\eea
Normalization requires $2\geq 1-(|E_k|^2-|F_k|^2)=2 |F_k|^2\geq 0$, and together with the renormalizability condition, we see that the energy density $\rho_{0m}$ is finite and definite positive.   Moreover,  we find  $   \frac{\left< T_{ii} \right>_{ren}}{a^2} \sim 0 \ , $ and hence the pressure obeys the cold matter equation of state.  On the other hand, for sufficiently early times in the evolution, $t << m^{-1}$, we have 
\bea
 \left<T_{00} \right>_{ren} \sim \frac{\rho_{0r}}{a^4} \ , \hspace{0.5cm}   \rho_{0r} =  \frac{1}{ \pi^2} \int_0^{\infty} dk k^3\left[1-(|E_k|^2-|F_k|^2)\right] \geq 0\ , 
\eea
and additionally $ \frac{\left<T_{ii} \right>_{ren}}{a^2} \sim \frac{1}{3} \left< T_{00} \right>_{ren} \label{Tii2/3} \ . $
 From this, we see that $p \sim \rho/3$, in agreement with the assumptions of classical cosmology for the radiation.

We notice that these are the leading contributions to the energy density and pressure when considering late and early times in the cosmological evolution, irrespectively of the particular choice for the vacuum state. Higher order adiabatic contributions shall provide quantum  corrections to all these well-known classical results, where different choices of the vacuum shall lead to different physical predictions.

\section{Equivalence to DeWitt-Schwinger Point-Splitting}

The adiabatic program is equivalent to the DeWitt-Schwinger point splitting technique.  Here we shall simply give a sketch of the proof, the details can be seen in [\refcite{papers}]. The key idea is to notice that two different methods to compute the renormalized stress-energy tensor can differ, at most, by a linear combination of conserved local curvature tensors, up to 4th orders in the derivatives of the metric. In other words,
\bea
  \langle T_{\mu \nu} \rangle^{Ad}_{ren} - \langle T_{\mu\nu} \rangle^{DS}_{ren}=   c_1 {^{(1)}}H_{\mu\nu} + c_2 {^{(2)}}H_{\mu\nu} + c_3 m^2 G_{\mu \nu}+ c_4 m^4g_{\mu\nu}  \ ,   \label{ambiguity}
 \eea
 where $g_{\mu\nu}$ is the spacetime metric, $G_{\mu \nu}$ is the Einstein tensor, and $^{(1)}H_{\mu\nu}$, $^{(2)}H_{\mu\nu}$ can be obtained by functionally differentiating the quadratic curvature Lagrangians $R^2$ and $R_{\mu\nu}R^{\mu\nu}$. The task is then reduced to determine the free coefficients. 

The DeWitt-Schwinger approach deals with the asymptotic expansion of the Green's function $S_{DS}(x, x')$, defined by
\bea
\left(i{ \gamma}^{a}e_a^{\mu}\nabla_{\mu}^x- m\right)S_{DS}(x, x')=|g(x)|^{-1/2}\delta(x-x\, ') \ .  \label{linearcombination}
\eea
For a Dirac spinor, $\langle T_{\mu}^{\mu} (x)\rangle_{ren} = m \langle \bar \psi \psi(x) \rangle_{ren}$ holds. Thus, the strategy to determine the free coefficients in (\ref{ambiguity}) follows by analyzing the two-point function.

The two-point function $^{(4)}\langle \bar \psi \psi (x) \rangle_{Ad}$ can be computed using the adiabatic method by descomposing the field in Fourier modes, expanding them adiabatically, and retaining contributions in the asymptotic / adiabatic structure up to four derivatives of the metric. In order to compare now the result with DeWitt-Schwinger, one follows an approach introduced by  Bunch and Parker  [\refcite{bunch-parker}]. It is an  alternative asymptotic expansion of the two-point function in momentum space, which was shown to be equivalent to DeWitt-Schwinger point-splitting. It was proposed to aim at extending to curved space the standard momentum-space methods of perturbation theory for interacting  fields in Minkowski space.  This way the standard Minkowskian propagator of a scalar free field in momentum space $(-k^2 +m^2)^{-1}$ is replaced by a series expansion. The Fourier transform leading to local-momentum space is crucially performed with respect to  Riemann normal coordinates $y^{\mu}$ around  a given point $x'$. In contrast to  adiabatic regularization, the method is valid for an arbitrary space-time. The method works directly with the two-point functions, which are regarded as the basic buildings blocks of the renormalization process,
\bea
tr S_{DS}(x, x')\sim \int \frac{d^3 \vec k}{(2\pi)^4} e^{ik \cdot y}\sum_n A_n(y)\left[-\frac{\partial}{\partial m^2}\right]^{(n)} \frac{1}{-k^2+m^2}\, .
\eea
This is a geometric asymptotic series, where $A_n(y)$ contains $2n$ derivatives of the metric. A detailed analysis can be carried out and the result is that
\bea
\lim_{x' \rightarrow x} Tr \ {^{(4)}}S_{DS}(x,x') =   ^{(4)}\langle \bar \psi \psi (x) \rangle_{Ad} \, .
\eea
Turning back to the trace of the stress-energy tensor in (\ref{ambiguity}), we see that all the coefficients equal to zero. Thus,  $\left<T_{\mu\nu}(x) \right>_{ren}^{Ad}=\left<T_{\mu\nu}(x) \right>_{ren}^{DS}$.

\section{Conclusions and final comments}

Adiabatic renormalization provides explicit expressions for the renormalized stress-energy tensor in expanding universes for spin-$1/2$ fields.  The result of the renormalized expectation values for the stress-energy tensor  is shown to be equivalent to the standard DeWitt-Schwinger point-splitting approach. Adiabatic, however, turns out to be much more convenient to apply when dealing with homogeneous spaces in cosmology, since the involved numerical computations are more efficient.

\section*{Acknowledgments}

This work is supported by Grants No. FIS2014-57387-C3-1, and No. CPANPHY-1205388. A.D. is supported by the Spanish Ministry of Education Ph.D. fellowship FPU13/04948.

\end{document}